\documentclass[aps,twocolumn,prd]{revtex4}
\usepackage{epsfig}
\usepackage{psfrag}

\input epsf

\def\be{\begin{equation}}
\def\ee{\end{equation}}
\def\bea{\begin{eqnarray}}
\def\eea{\end{eqnarray}}

\def\cmm2{{\,\rm cm^{-2}}}
\def\cm2{{\,{\rm cm}^2}}
\def\cmm3{{\,{\rm cm}^{-3}}}
\def\gcmm3{{\,{\rm g\,cm^{-3}}}}

\def\mpl{{m_{\rm Pl}}}

\def\fun#1#2{\lower3.6pt\vbox{\baselineskip0pt\lineskip.9pt
  \ialign{$\mathsurround=0pt#1\hfil##\hfil$\crcr#2\crcr\sim\crcr}}}

\def\la{\mathrel{\mathpalette\fun <}}

\def\fun#1#2{\lower3.6pt\vbox{\baselineskip0pt\lineskip.9pt
  \ialign{$\mathsurround=0pt#1\hfil##\hfil$\crcr#2\crcr\sim\crcr}}}

\def\mpl{{\rm m}_{\rm pl}}
\def\dw{\Delta w}

\def\om{\Omega_M}
\def\gL{\Lambda}
\def\wcmb{\bar w_{\rm cmb}}
\def\wsn{\bar w_{\rm sn}}

\def\ltsima{$\; \buildrel < \over \sim \;$}
\def\simlt{\lower.5ex\hbox{\ltsima}}
\def\gtsima{$\; \buildrel > \over \sim \;$}
\def\simgt{\lower.5ex\hbox{\gtsima}}

\begin{document}
\bibliographystyle{prsty}
\title{Testing for a Super-Acceleration Phase of the Universe}
\author{Manoj Kaplinghat}
\email{mkapling@uci.edu}
\affiliation{Department of Physics and Astronomy,\\
University of California, Irvine, California 92612, USA}
\author{Sarah Bridle}
\email{sarah@ast.cam.ac.uk}
\affiliation{Institute of Astronomy, University of Cambridge\\
Madingley Road, Cambridge, CB3 0HA, UK}
\date{\today}

\begin{abstract}
We propose a method to probe the phenomenological nature of dark
energy which makes no assumptions about the evolution of its energy
density. We exemplify this method with a test for a super-acceleration
phase of the universe i.e., a phase when the
dark energy density grows as the universe expands. We show how
such a phase can be detected by combining SNIa (SNAP-like) and CMB
(Planck) data without making any assumptions about the evolution of
the dark energy equation of state, or about the value of the matter
density parameter.

\end{abstract}
  \maketitle

\section{Introduction\label{sec:intro}}
Observations of distant Type Ia supernovae imply a presently
accelerating universe \cite{riess98, perlmutter99}. In the standard
cosmological model this is accommodated by introducing ``dark energy'',
a component which interacts with the rest of the universe only
gravitationally and has a  significantly negative pressure which leads
to the acceleration of the universe. 

In order to accelerate the universe, the dark energy component must
have an energy density which decreases (if at all) much more slowly
than matter density as the universe expands. Current data favor a dark
energy density which is almost constant or increasing with time
\cite{caldwell99,schuecker02,tonry03,knop03,choudhury03,alam03} and 
exciting results can be expected in the future
\cite{weller01,frieman02,linder03,kratochvil03}. We label the phase 
when the dark energy density is increasing with the expansion of the
universe as super-acceleration.

In this paper we discuss a method to probe phenomenological properties 
of dark energy without any assumptions about the evolution of dark 
energy density. We use this method to formulate a test to ascertain if 
the universe evolved through a super-acceleration phase by  combining 
SNIa and CMB experiments. We quote our results in terms of two variables 
$\wsn$ and $\wcmb$ which are the {\em best fit} constant dark energy 
equation of state parameters for the SNIa and CMB experiments
respectively.  Our method does not rely on the equation of state
parameter being constant. In fact, we do not need to assume anything
about the evolution of the dark energy equation of state, 
or about the value of the matter density parameter.

We will outline our method in the context of the specific example 
of testing for super-acceleration using SNIa and CMB experiments. 
We will end with a discussion of how our method can be generalized to 
other questions one might ask about dark energy using SNIa, CMB and 
other experiments. Questions of the nature, ``did the
dark energy density increase with cosmic expansion?'' are perhaps
easier to answer than some questions previously asked in the literature.
And yet they have the potential to revolutionize
our thinking about dark energy. We stress that answering such
questions does not necessarily require measuring the complete dark
energy density evolution; the key point is that we are asking for
less information. 

In a nutshell, to apply this method one needs to fit a constant
equation of state $\bar{w}$ to each of SNIa and CMB observations and plot
conditional probabilities in the $\bar{w}$ versus $\Omega_M$ plane
(e.g., Fig.~1). The probabilities are conditional in the sense that
they are normalized at each value of $\Omega_M$.
If the upper contours cross below $\bar{w}=-1$ then super-acceleration
has been detected. We stress that the power to make such robust
statements comes from combining different measures (SNIa and CMB)
which have complementary dependencies on the dark energy equation of
state and matter density.

In Section \ref{theory} we start with a discussion of dark energy and
define super-acceleration rigorously. We then outline the general
characteristics of models which could describe a super-acceleration
phase. 
Given that dark energy is a total mystery even if it never went
through a super-acceleration phase, our  point of view is that
super-acceleration is a question to be observationally settled. 
It is satisfying to note that this will be possible with upcoming
experiments. We explain our evolution-independent method in
Section \ref{sec:method}. We then proceed to forecast the  possibilities
with regard to measuring a super-acceleration phase from  future
missions in Section \ref{sec:forecasts}. We end with a discussion of  how
the method can be generalized. 

\section{Dark Energy and Super-Acceleration \label{theory}}
In this section we will consider dark energy in a general setting (without
reference to a theory of gravity) and define it in terms of its 
phenomenological properties. This will allow us to motivate 
super-acceleration as a part of phenomenologically viable parameter 
space. Finally, we will consider some models where a super-acceleration 
phase might arise. These models are discussed with an aim to better 
understand the requirements that super-acceleration imposes.

\subsection{Generalized Equation-of-state\label{sec:w}}
The pressure of the dark energy (DE) $p_X$ is parameterized in terms of
the equation of state $w_X$ which relates it to the physical energy
density $\rho_X$ as $p_X = w_X \rho_X$. In terms of $w_X$ the
evolution of the DE density with scale factor $a$ is  
$\rho_X(a) \propto \exp(-3 \int^a d\ln a' (1+w_X(a'))$. If 
$w_X < -1$, then the DE density is actually increasing as the universe 
expands, which seems counter-intuitive. In the context of a scalar
field ($\phi$) with canonical kinetic ($\dot \phi^2/2 > 0$) and
potential ($V(\phi) > 0$) terms, $w_X < -1$ is unphysical. This is
easily seen by rewriting  $w_X$ as  $1+w_X = 2\dot \phi^2/[\dot \phi^2
  + 2 V(\phi)]$ which is positive if $V(\phi) > 0$.

The SNIa observations provide us with a magnitude--redshift
relationship. If the SNIa can be standardized as sources with known
luminosity, then they measure the luminosity distance $d_L(z)$ in units
of $1/H_0$ where $H_0$ is the present expansion rate of the
universe. Within the context of homogeneous and  isotropic universes,
we can then write  
\be 
D(z)\equiv {H_0 d_L(z) \over (1+z)}= \int_0^z {dz \over h(z)}\,, 
\label{Dz}
\ee
where $h(z)=(\dot a /a)/H_0$ is the expansion rate of the universe
normalized to unity today and $z=1/a-1$ is the redshift. Acceleration
refers to the condition $\ddot a > 0$. We can parameterize the
expansion rate as  
\be
h^2(z)=(1+z)^3 \left[\Omega_M +
\Omega_X \exp( 3\int_0^z d\xi \, w_X(\xi)/(1+\xi))\right].
\label{Hz}
\ee
This is completely general and it does not tie us down to a theory of
gravity (see \cite{tegmark01,carroll01,linder02,dvali03,linder03} for
examples of studying cosmology in this spirit). Any $h(z)$ can be written
in terms of $w_X(z)$ as above. Eq. \ref{Hz} is a phenomenological fit;
we need a theory (for gravity, dark matter  and dark energy) to
interpret $\Omega_M$ and $w_X(z)$.

We can differentiate Eq. \ref{Hz} to derive the acceleration of the
scale factor. This yields
\be
2 \ddot a / a = -(1 + 3 w_X(z)) h^2(z) + 
3 w_X(z) (1+z)^3 \Omega_M \,.
\label{addot}
\ee
The universe is accelerating if $w_X(0) < -1/[3(1-\Omega_M)]$. 
Also, for  $-1/[3(1-\Omega_M)] > w_X(0) > -1/(1-\Omega_M)$ we
have $\ddot a(0) > 0$ but $\dot h(0) < 0$. 

\subsection{Super-Acceleration\label{sec:sa}}
We will refer to an expansion phase with $w_X(z) < -1$ as
super-acceleration \footnote{Models with similar phenomenology were
  first considered in the context of (super-exponential) inflation in
  higher-dimensional gravity theories with non-canonical derivative
  terms \cite{pollock88}.}.
Super-acceleration implies that the dark energy contribution to the
expansion rate is increasing with time and this leads to interesting
scenarios for the ultimate fate of the universe
\cite{mcinnes01,caldwell03}. In particular, it is 
possible that the universe could end in a ``Big-Rip'', with the time
scale for such catastrophic events set by $1/H_0$. In order to draw
such conclusions, the future evolution of $w_X(z)$ must be known. 

We are interested in the issue of whether future observations will be
able to ascertain that the universe went through a super-acceleration
phase. The main motivation for thinking about super-acceleration is
the simple fact about dark  energy -- we don't know
what it is. Given our current understanding, even if we know what
$w_X(z)$ exactly is, dark energy will still be a mystery. 

One of the troubling things about super-acceleration is that it seems
to violate causality. The magnitude of the pressure is 
larger than the energy density and hence one might expect to transmit
information faster than the speed of light. While this is a 
possibility, it is by no means guaranteed by having 
$|p_X| > \rho_X = 3 \mpl^2 H_0^2 [h^2(z) - (1+z)^3\Omega_M]$. 
$w_X(z)$ determines the global evolution of $\rho_X$; it does not tell
us how small local perturbations propagate. 

\subsection{Modeling Super-Acceleration\label{sec:models}}
Given the repercussions and the unintuitive nature of $w_X(z) < -1$
systems, it is of interest to investigate models which give rise
to a super-acceleration measurement. 

\begin{enumerate}
\item One possibility is that the universe is accelerating due
to a cosmological constant (or perhaps some dynamical scalar field),
but the effective $w_X$ at some epoch goes below -1 because of new
physics we are unaware of. An example of new physics \cite{kaloper03}
could be the coupling of photons to axions (not the QCD axion)
\cite{csaki01} which has been proposed to explain the dimming of
supernovae without the need for an accelerating universe.   

\item A second (troubling) possibility is that the effective $w_X$ is
less than -1 because of systematic effects. This is similar to (1)
except it is not new physics but some systematic effect that results
in the ``apparent'' super-acceleration.

\item $w_X < -1$ is a possibility if one postulates that dark energy
  is a scalar field with non-canonical kinetic terms
  \cite{caldwell99,picon99,chiba99,frampton02}. One must be
  careful about instabilities in the theory in this case \cite{carroll03}.  
     
\item Another class of models appeals to a modification of gravity on
  scales $\la 1/H_0$ to obtain $w_X(z) < -1$. Like the other models in
  this section, this class of models does not solve the ``why now?''
  problem since the scale $H_0$ is put in by hand to make sure that
  gravity is only modified today.  There are severe constraints on
  modified gravity theories in the early universe
  \cite{carroll01}. The main impetus for considering modified gravity
  theories is provided by brane-world models  
  wherein it might be possible to get super-acceleration 
  \cite{sahni02,dvali03}.   

\end{enumerate}

\section{Detecting Super-Acceleration\label{sec:method}} 

In this section we lay out the method by which future experiments can
unambiguously detect a super-acceleration phase, independent of the
functional form of $w(z)$. First we show that if a constant equation
of state is fitted to SNIa observations and a result less than $-1$ 
is obtained then either the true equation of state must at some point
dip below $-1$, or the assumed matter density is incorrect.
We then show the same thing for CMB observations.
Finally we show how SNIa and CMB information together can overcome the 
need to assume a value for the matter density.
Throughout we assume a flat universe.

\subsection{Information from Supernova observations}

Given a set of SNIa observations our method uses conventional 
$\chi^2$ fitting to obtain the best fitting constant equation
of state for each $\Omega_M$ (see next section).
The aim of the following paragraphs is to show the significance 
of this fitted constant equation of state, given that the actual
equation of state is of unknown functional form.

A SNIa experiment measures the luminosity distance 
$D_E$ out to a 
certain set of redshifts. 
As usual, we define:
\be
D_E(z)={1 \over \sqrt{\Omega_M}} \int_0^z 
{dz \over \sqrt{(1+z)^3+r_E(z;\Omega_M)}}\,,
\label{DEdef}
\ee
where $r_E(z;\Omega_M)\equiv (\Omega_X/\Omega_M)\exp[3\int_0^z 
dz'(1+w_X(z'))/(1+z')]$. The subscript $E$ in Eq. \ref{DEdef}
explicitly denotes that this is the quantity measured by an
experiment. We are further assuming that the experimental magnitude
redshift relation is an unbiased estimator of the actual luminosity
distance.  

The next step in our method is to fit the distance $D_E(z)$ with a
distance $D_F(z,\Omega_M,\dw)$ which assumes a constant
equation-of-state,    
$w_X(z)=\wsn=-1+\dw$. 
Thus $D_F$ is has the same form as $D_E$ but with $r_E$ replaced by 
$r_F(z;\Omega_M)=(\Omega_X/\Omega_M)(1+z)^{3\dw}$. 
The best-fit $\dw$ is found by maximizing the likelihood assuming 
gaussian errors (appropriate for SNAP) for the measured distances. 

We give a more rigorous derivation in the following
paragraphs, but to gain a qualitative understanding first consider the  
weighting function approach of Saini et. al. \cite{saini03}.
They showed that to a good approximation a constant fitted equation
of state from supernovae observations is simply a weighted
average over redshift of the true equation of state, where the 
weighting function is always positive.
From this it is clear that if the constant fitted equation of 
state is less than $-1$ then the true equation of state must
also go below $-1$ for some redshift interval(s).
The above is only true if the correct matter density has been
assumed. 

We now have the task of computing the best-fit
$\dw$ for a given $D_E(z)$ at a fixed $\Omega_M$. This is found by
minimizing the $\chi^2$ at fixed $\Omega_M$ with respect to $\dw$. The
resulting equation does not have a simple solution. Our aim is to
figure out when the best-fit $\dw < 0$. To achieve this, we replace
$D_F$ in the minimization equation by its a first order Taylor
expansion in $\dw$, i.e.,
\be
D_F(z,\Omega_M,\dw) \simeq D_\gL (z,\Omega_M) + 
\dw\, G_\gL (z,\Omega_M)\,.
\label{DFapprox}
\ee
$D_\gL$ is defined the same way as $D_F$ but with $\dw=0$, i.e.,
$D_\gL (z,\Omega_M)$ is the distance in $\gL$CDM cosmology. We define
$G_F = d D_F/d\dw$ and the function $G_\gL(z,\om)=G_F(z,\om,\dw=0)$. For
$\Omega_M > 0.1$ and $0<z<1.7$, the error in the  approximation given
by Eq. \ref{DFapprox} is smaller than 3\% for $-0.5 < \dw < 0.5$. We  
note that Saini et. al. \cite{saini03} have shown that this kind of
linear approximation works extremely well even when $\dw$ is not
constant (in which case we need to take a functional derivative with
respect to $\dw(z)$.) 

We note three facts. One, we are only \emph{approximating the fit}
(to ascertain whether $\dw < 0$) and not making any assumptions about
the actual $w_X(z)$. 

Two, the error in approximating $D_F$ according
to Eq. \ref{DFapprox} turns out to be unimportant. At the end of this
section, we show explicitly that the only effect of this approximation
is to make our results (on detecting super-acceleration) somewhat
conservative. 

Three, we could just as well use $D_F=D_\gL+\dw G_\gL$
as our {\em exact} fit and it would not change any of our
arguments. We chose not to go this route because the above fit is
unintuitive and has unphysical behavior for large $\dw$. 

Applying the approximation in Eq. \ref{DFapprox} to the equation
resulting from the minimization procedure, we get the following
implicit equation for the best-fit $\dw$:
\be
\dw=\frac{ \Sigma_i \, \left[D_E(z_i)-D_\gL(z_i,\om)\right]
G_F(z_i,\om,\dw)/\sigma^2(z_i) }
{ \Sigma_i \, G_F(z_i,\om,\dw)G_\gL(z_i,\om)/\sigma^2(z_i)} .
\label{dw}
\ee
Both $G_F$ and $G_\gL$ are negative for all $z$. What
Eq. \ref{dw} says is that if the best fit $\dw <0$, then 
$D_E(z_i) > D_\gL(z_i,\Omega_M)$ for a finite number of $z_i$ or
$\Omega_M$ is wrong. 
This in turn implies that if $\om$ is the true value, then $w_X(z)$
must be less than -1 for some redshift range.

A point of note here is that for each $\Omega_M$ we have a
corresponding best-fit $\dw(\Omega_M)$. We are not minimizing with
respect to both $\Omega_M$ and $\dw$. The $\Omega_M$ resulting from a
joint minimization need not be the true matter density and in fact it
could be that the best-fit  $\dw < -1$ even though the underlying
$w_X(z)$ never enters a super-acceleration phase \cite{maor01}.

\begin{figure}
\psfrag{weff}{${\bar w}_{\rm sn,cmb}$}
\centerline{\psfig{file = 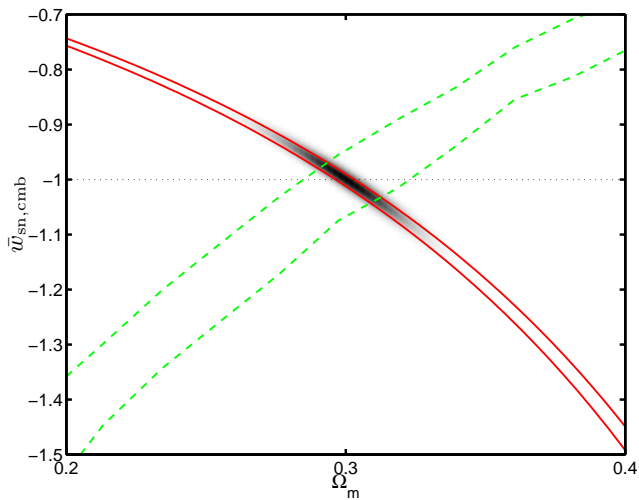,width=8.5cm}}
\caption{
  The shading shows the two-dimensional probability distribution
  from SNAP-type supernova data in the $\wsn$--$\om$ plane. The solid
  lines show 68 per cent limits at each $\om$ from the same simulated
  data. The dashed lines show 68 per cent limits at each $\om$
  from a Planck-type simulation, marginalized over 5 other
  cosmological parameters. The input model has $\om=0.3$
  and $w(z)=-1$ (a constant).  
}
\label{Fig:weffom_sncmb_wminus1}
\end{figure}

How are our results affected by 
the small error in approximating the fit luminosity distance
$D_F$ by a first order Taylor expansion? To answer
this question let us write 
\begin{eqnarray}
D_F(z,\om,\dw) & = & (D_\gL(z,\om)+\dw\, G_\gL(z,\om)) \nonumber\\
&\times & (1 -\epsilon(z,\om,\dw))\, .
\end{eqnarray}
For the interesting parameter space, $\epsilon(z,\om,\dw)$ is smaller
than about 0.03. Further, for all practical purposes, this
approximation overestimates $D_F$. One can then
show that the effect of this is to make $\dw$ smaller. Thus the
conclusions drawn from Eq. \ref{dw} are conservative in the sense
that $\dw$ is more negative than what we would find from Eq.~\ref{dw}. 

An interesting exercise is to fit the supernova distances with 
$D_F(z,\om,\dw)= D_\gL(z,\om)+\dw\, G_\gL(z,\om)$ {\em exactly}. We
can then run through the same arguments as before and arrive at the
same conclusion. In particular, we computed the new contours with
$w_{\rm sn}\equiv -1+\dw$ and the quantitative results changed very  
little as expected. 

\subsection{Low redshift supernova calibration}

In the discussion above, we have assumed that we can measure luminosity
distance $D_L$ as a function of redshift from supernovae. However in
practice we can only measure magnitudes $m$ 
\begin{equation}
m = \mathcal{M} + 5\log D_L(z,\Omega_m, w(z))\,,
\end{equation}
where the calibration factor $\mathcal{M}$ is unknown unless we have a
very large number of low redshift supernovae. 

We expect a few hundred supernovae from SNFactory
\cite{aldering02} and the low redshift Carnegie Supernova Program
\cite{phillips04} and that will enable calibration of the
luminosity distances to around 0.5 per cent. If the $\sim$6000
supernovae that should be seen by GAIA \cite{belokurov03} were
all followed up then this could reduce the uncertainty to 0.2 per
cent. 

This additional uncertainty will reduce our ability to determine if
super-acceleration occurred. We take the conservative (and not fully
optimized) approach
of assuming the worst possible value for the calibration parameter and
feeding it through the analysis described above. If the luminosity
distance calibration were actually a factor of 1.015 (3$\times$0.5 per
cent) higher than estimated from the low redshift sample, then the
supernova contours would shift from the solid lines in
Fig.~\ref{calibfaff} to the dotted lines. This factor corresponds to
the true calibration factor being about three $\sigma$ away from that
estimated by the upcoming experiments. If there were significantly
more supernovae followed up, then this would improve the picture. For
example, we get the dashed contour with a factor of 1.006.

\subsection{Information from CMB observations}

The CMB is sensitive to the dark energy mainly through its effect
on the angular diameter distance to the CMB, and therefore the 
positions of the acoustic peaks \cite{huey98}. 
It also has an effect on the largest scales since dark energy 
changes 
the Integrated Sachs Wolfe effect,
through
both the change in the expansion rate and clustering.
Here we propose to ignore the CMB
information on large scales so that there is no need to make
assumptions about the clustering of dark energy.

In the case of the CMB, the approximation given in Eq. \ref{DFapprox}
does not work well. However we do not need the approximation since the
CMB measures the angular diameter distance out to just one redshift,
that of the last scattering surface. The equation to solve for the CMB
is $D_F(\infty,\om,\dw)=D_E(\infty)$ (neglecting the small error on
$D_E$). Although the relation between $\dw$ and $D_E(z)$ is
non-trivial here, the argument is simple. For the true $\om$ if the
relation $D_F(z,\om,\dw)=D_E(z)$ implies that 
$\wcmb \equiv -1 + \dw <-1$ then a
super-acceleration phase must happen over a finite range of
redshift. This is simply a consequence of the fact that
$D_F(z,\om,\dw)$ is a monotonically decreasing function of $\dw$.  

Qualitatively, when a constant 
equation of state is fitted to CMB data it measures roughly
a weighted average of the true $w_X(z)$ where the weighting function 
is always positive \cite{saini03}.
Therefore if the fitted $\wcmb$ gives a value less than $-1$ then
the true equation of state must itself dip below $-1$.
As for the SN1a argument,
this is only true if the value of $\om$ has been guessed correctly.

\subsection{Combination}

\begin{figure}
\psfrag{weff}{${\bar w}_{\rm sn,cmb}$}
\centerline{\psfig{file = 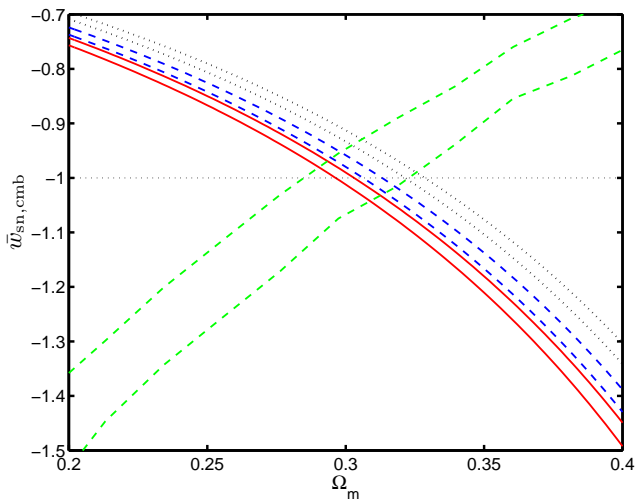,width=8.5cm}}
\caption{
The contours in this figure are computed the same way as the contours
in Fig.~\ref{Fig:weffom_sncmb_wminus1}. The solid supernova contour
assumes that the calibration error is negligible. The dashed lines are
68 \% limits assuming a worst case calibration parameter estimated
from 6000 low redshift supernovae. The dotted line is the same as the
dashed but for 300 low redshift supernovae. 
\label{calibfaff}
}
\end{figure}

As discussed above, using SNIa alone it is not possible to
uncover the nature of dark energy simultaneously with estimating
the matter content $\Omega_M$.
This is also true for CMB data alone.
Therefore we need to combine information from SNIa and CMB.
Given observed data we can determine the
limits on $\bar{w}$ at each $\om$ for both SNIa and CMB and plot this
conditional likelihood in the $\bar{w}$--$\om$ plane.  
In general the best-fit $\bar{w}$ at some $\om$
derived from SNIa will be different from that derived from CMB
since they probe the equation of state at different redshifts. 
For illustration we use contours which encapsulate 68\% of
the likelihood throughout, however the method can be carried
out for whatever confidence level is required.

The key point of our argument is that if, for
every value of $\om$, the contour for either $\wcmb$ or $\wsn$ falls  
below -1, then it must be that dark energy went through a phase of
super-acceleration for a finite period of time. We explain how this
follows from the previous sections by discussing two examples. 

Consider the example in Fig.~\ref{Fig:weffom_sncmb_wminus1}.
Suppose that we do not know the true value of $\om$.
We get around this problem by considering each possible value
for $\om$ in turn.
For each $\om$ value we assess whether, if this were indeed the
true value of $\om$, we would conclude that the universe underwent
super-acceleration.
If for all values of $\om$ we reach the same conclusion, then we 
can give a definitive answer, independent of the value of $\om$.

As an example let us first consider the value $\om=0.2$.
In this case $\wsn>-1$ and $\wcmb<-1$. 
If this were the true value of $\om$ then we would have to conclude
that at some point in the history of the universe, $w(z)<-1$.
(In fact since the CMB probes the value of $w(z)$ at higher
redshifts than SNIa then this would imply that $w$ was once
less than $-1$ and has since risen above $-1$, but these details
are not important for our argument.)
We would reach this same conclusion for all matter densities
with $\om\simlt0.28$.
Now consider the value $\om=0.4$. 
If this were the true value of $\om$ then we would again conclude
that for some redshifts $w(z)<-1$, since $\wsn<-1$.
We would reach the same conclusion for $\om\simgt0.305$.
However, for $\om=0.3$ then \emph{both} data sets allow $\bar{w}>-1$,
therefore we cannot be sure whether or not there was super-acceleration.
Our conclusion from Fig.~\ref{Fig:weffom_sncmb_wminus1} is that
either $0.28 \simlt \om\simlt 0.305$ or there was super-acceleration.

Now consider future observations providing the solid (SNIa) and
dashed (CMB) lines in Fig.~\ref{Fig:wfiddle}.
Running through the same arguments as above, we conclude that,
no matter what the value of $\om$, there must have been
super-acceleration. This can be seen at a glance by the fact that the
upper contours cross below $-1$.

The method we have just outlined determines if the universe underwent
a super-acceleration phase without any assumptions about the evolution
of dark energy density. 
It is clear from the above discussion that one cannot test for a
super-acceleration phase if the deviation from $w_X=-1$ is arbitrarily
tiny. How well can we do with future experiments? We turn to this
question next.

\section{Forecasts\label{sec:forecasts}}

In order to make a prediction about how well future experiments will
do, we need to generate mock data ($D_E$) which in turn 
requires that we make some assumptions about the evolution of
$w_X(z)$. This is not a major hurdle (as we show later), mostly
because we interested in getting a rough sense of what future
experiments can achieve. In the next section, we will discuss the
procedure used to generate mock data assuming a particular $w_X(z)$
function.  We generate data relevant for SNAP-like and Planck
experiments.   

\subsection{SNIa data\label{sec:forecasts.snia}}

We assume that the luminosity distance will be measured to an accuracy
of 1.4 per cent in each of 50 redshift bins 
evenly spaced between redshift 0.1 and 2. First we simulate data
from a model with $w(z)=-1$ and $\om=0.3$.

On fitting the two parameters, $\wsn$ and $\om$ we obtain a tight
constraint on the two parameters indicated 
by the shading in Fig.~\ref{Fig:weffom_sncmb_wminus1}.

For our method the quantity of interest is not the two-dimensional
probability (as shown by the shaded contour) but the confidence limits
on $\wsn$ calculated for a given $\om$. These limits are shown by the
solid lines on Fig.~\ref{Fig:weffom_sncmb_wminus1} and as expected the
contours extend to $\wsn< -1$ in the direction of increasing $\om$.

\subsection{CMB data\label{sec:forecasts.cmb}}

We generate CMB temperature anisotropy (TT) data using 
the parameters for the Planck HFI 353 GHz channel: sensitivity
$\Delta T/T = 14.4\times 10^{-6}$,  solid angle per pixel 
$4 \pi/ 5.94 \times 10^6$  (where $5.94 \times 10^6$ is the number of
pixels on the sky) and 5 arcmin FWHM beam and assume 100 per cent of
the sky can be used. To speed up the computation we bin the
observations with  $\Delta \ell=10$ and go up to 
$\ell_{\rm max}=2500$. 
We also marginalize over the Hubble constant, the baryon content, and
the amplitude and spectral index of the primordial scalar
fluctuations. 
We ignore tensor modes and do not use CMB polarization
information, although in principle these could both be included
in the analysis.

The CMB data essentially constrain the angular diameter
distance to last scattering $z\sim 1100$ and therefore there is a
perfect degeneracy between $\wcmb$ and $\om$. Including the large scale
ISW effect (the decay of gravitational potential due to dark energy
domination) or lensing on small scales would break this degeneracy to
some extent. However, a reliable estimation of the degeneracy
breaking would require some assumptions about the clustering of dark
energy. In light of the uncertain nature of the models of dark energy
with $w_X(z) < -1$, we have decided to not use the lensing signal or
the $\ell < 50$ power spectrum (where the ISW effect is important). 
We again plot contours at the 68 per cent confidence limits for each
$\om$ value, although the more conventional contours at
constant two-dimensional probability look similar.

\subsection{Requirements on w(z) for detection}

\begin{figure}
\psfrag{weffsn}{${\bar w}_{\rm sn}$}
\psfrag{weffcmb}{${\bar w}_{\rm cmb}$}
\centerline{\psfig{file = 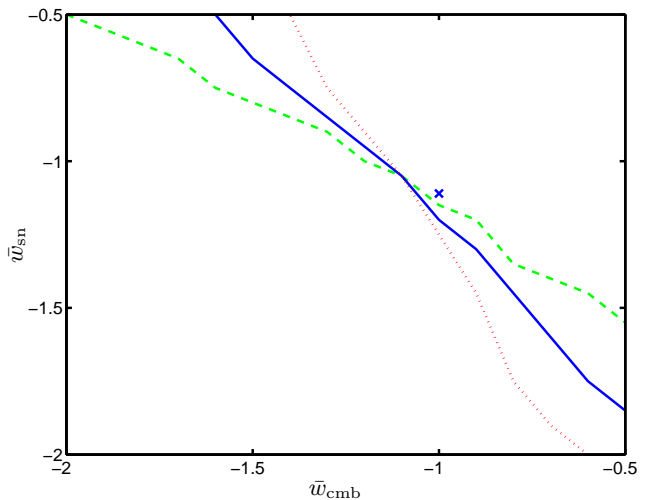,width=8.5cm}}
\caption{
The value of $\wsn$ below which it is possible to detect
super-acceleration, for a given $\wcmb$ and $\om$. 
The solid curve corresponds to a fiducial matter density of
$\om=0.3$, the dashed line corresponds to $\om=0.2$ and the 
dotted line corresponds to $\om=0.4$.
The cross shows the $\wsn$ at which super-acceleration can be detected
for an evolving equation of state with a jump in $w_x(z)$ at
redshift 0.3 for $\om=0.3$.
}
\label{Fig:wsn-wcmb}
\end{figure}

In the simplest approximation the shape of the contours in 
Fig.~\ref{Fig:weffom_sncmb_wminus1} is independent of the
shape of $w_X(z)$. 
For now we assume this to be true and assume the shape is 
the same as that for a constant $w_X$. We discuss the validity of this
approximation in the following subsection.

We produce plots similar to Fig.~\ref{Fig:weffom_sncmb_wminus1} 
but simulate data for models with various values for the
constant equation of state. We fix $\Omega_M=0.3$ throughout
and allow different values for the equation of state for the
CMB simulations and the SNIa simulations. This corresponds
to the case where the equation of state is varying with 
redshift and therefore the effective value is different for the
two probes.
For each assumed $\wcmb$ we calculate the $\wsn$ for which 
the upper CMB and SNIa 68\% contours cross below $-1$, and therefore
it is possible to detect super-acceleration.
This is shown by the solid line in~\ref{Fig:wsn-wcmb}.
For this value of $\wsn$ and lower it is possible to detect
super-acceleration. 
The dashed and dotted lines show the equivalent for 
$\Omega_M=0.2$ and $\Omega_M=0.4$.

This takes into account the error bars predicted for
SNAP-like and Planck-like experiments.
For example, if the true value of $\Omega_M$ is 0.3 then if
we measure $\wcmb=-1$ then $\wsn$ must
be less than $-1.3$ if we are to detect super-acceleration
from these future experiments.
Note that it is not necessary for us to assume a value for
$\Omega_M$ in order to detect super-acceleration; 
however the true value of $\Omega_M$ will affect how 
easy it would be to detect such a phase. 

In making Fig.~\ref{Fig:wsn-wcmb}, we have assumed that the
calibration error is negligible. Depending on the calibration error,
the curves in Fig.~\ref{Fig:wsn-wcmb} will shift down somewhat.  
The magnitude of this effect may be ascertained by looking at the
shift of the contours in Fig.~\ref{calibfaff}. 

\subsection{Insensitivity to assumed fiducial $w_X$\label{sec:forecasts.wx}}

\begin{figure*}
\psfrag{weff}{$\bar{w}_{\rm sn, cmb}$}
\psfig{file = 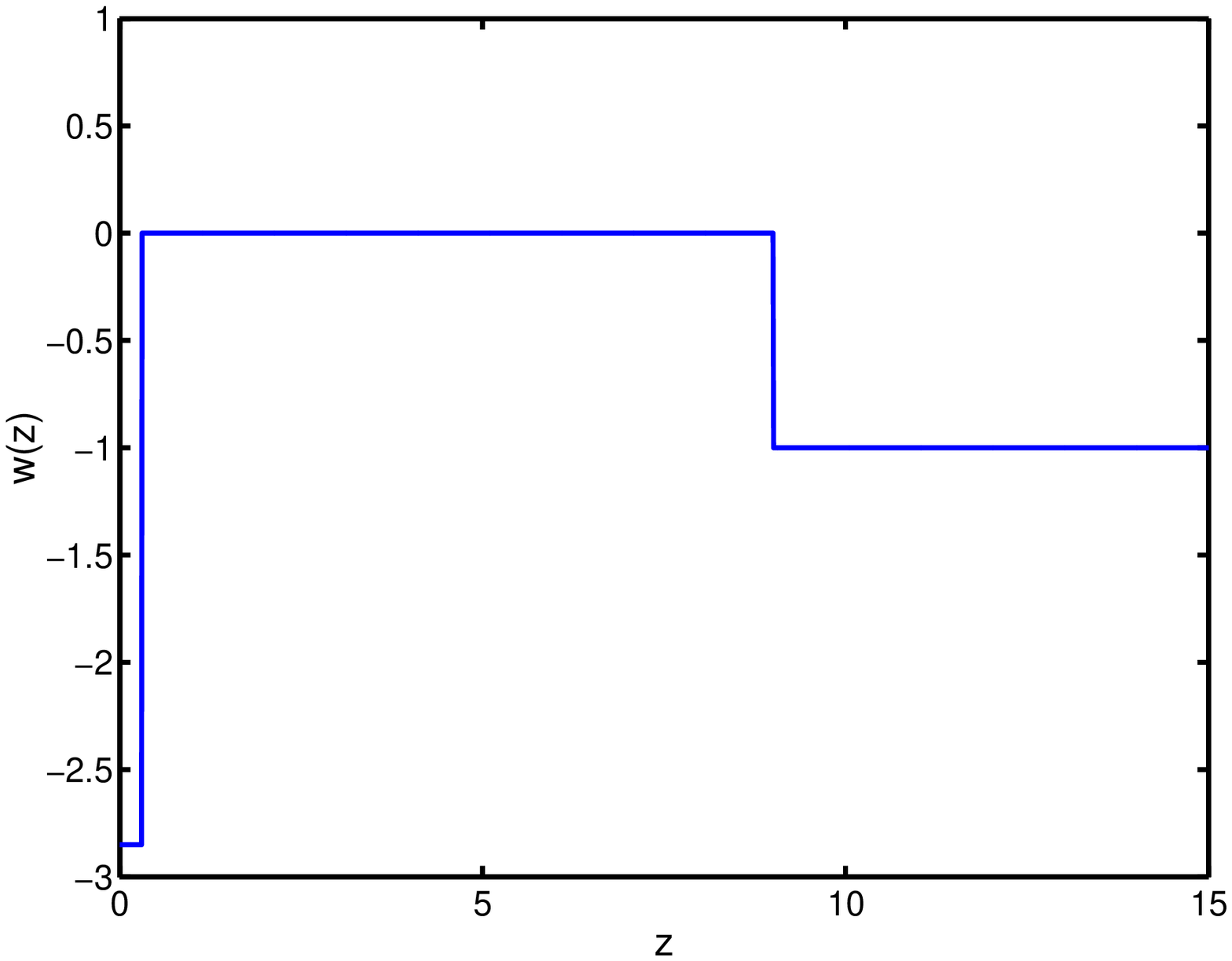,width=8.5cm}
\psfig{file = 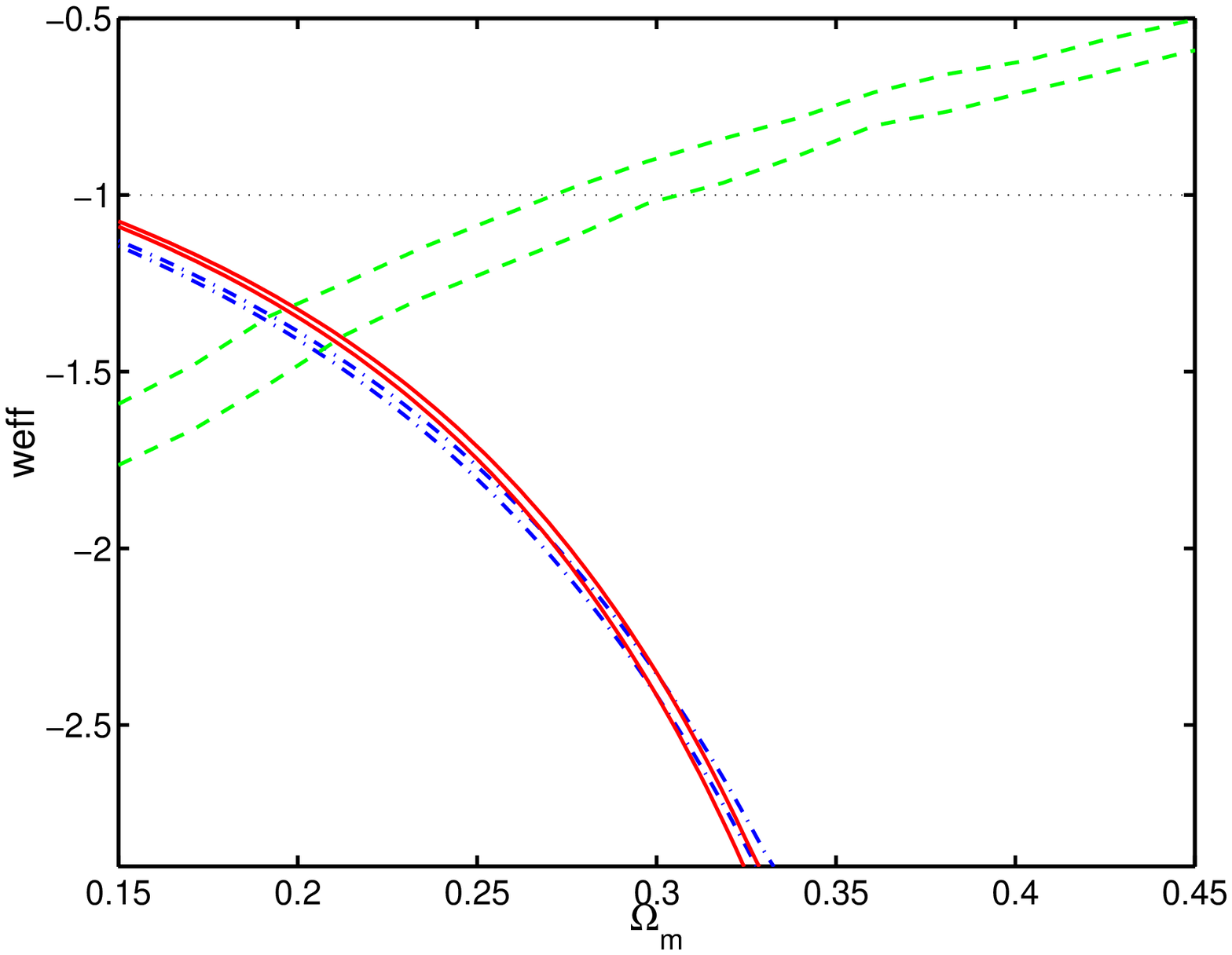,width=8.5cm}
\caption{
(a) The equation of state as a function of redshift for an extreme
strongly evolving model. A large jump occurs at $z=0.3$.
(b) Solid blue lines: 68 per cent confidence limits on
$\wsn$ for each $\om$ value from a SNAP-type
simulation with w(z) given by Fig.~3(a).
Dash-dot lines: the simulation has
$w(z)=-2.38$, the best fitting constant equation of state
model to the supernova data.
The solid and dash-dot lines are quite similar, despite the
fact that the $w_X$ evolution is so different in the two cases.
Since this w(z) was devised to give the same angular diameter
distance to the CMB as a $w=-1$ model, the CMB contours
(dashed lines) are unchanged from Fig.~1.
}
\label{Fig:wfiddle}
\end{figure*}
%

To calculate the lines in Fig.~\ref{Fig:wsn-wcmb} we assumed that
the shape of the contours (e.g., of Fig.~\ref{Fig:weffom_sncmb_wminus1})
are independent of the form of $w(z)$.
We take two approaches to assess the reliability of this approximation.
First we calculate the mathematical requirements for this 
approximation to be true.
Then we take an extreme example of a sharply changing $w(z)$ and
show how even in this case, our predictions in 
Fig.~\ref{Fig:wsn-wcmb} are not altered much.

To determine the mathematical requirements for our approximation 
let us assume we simulate mock data with a given value of matter density
denoted by $\om^t$ (``true matter density'') and a particular
evolution for the dark energy density. 
For this $D_E(z)$ we find the best fit $\bar \dw(\om)$ (as described
above in Section \ref{sec:method}). We are interested in the question
of how $\dw(\om)$ changes due to a change in $D_E(z)$ with the
constraint that $\dw(\om^t)$ is unchanged. Let us  consider the
following change $D_E(x) \rightarrow D_E(x) + \delta D_E(x)$. Then  
\bea
& & \dw(\om) = \bar \dw (\om) + \nonumber \\
& & \frac{\Sigma_i \, \delta D_E(z_i) G_F(z_i,\om^t) 
W_F(z_i,\om,\om^t)/\sigma^2(z_i) }
{\Sigma_i \, G_F(z_i,\om)G_\gL(z_i,\om)/\sigma^2(z_i)},
\label{DEchange}
\eea
where we have dropped the explicit dependence on $\dw$ in the right
hand side of equation. In the above Eq. \ref{DEchange},
\bea
& & W_F(z,\om,\om^t,\dw) = \frac{G_F(z,\om,\dw)}{G_F(z,\om^t,\dw)}-1 
\nonumber \\
& & \approx (\frac{\om}{\om^t}-1) \frac{d\ln G_F(z,\om,\dw)}
{d \ln \om} |_{\om^t}\,,
\label{WF}
\eea
where the last approximation is valid for small departures 
from $\om^t$. This is in fact the situation we
will be interested in. In this regime, $W_F(z,\om,\om^t)$ is sensibly 
constant. If $W_F(z)$ is truly constant, then $\dw(\om) = \bar
\dw(\om)$. Thus we have shown that a $\dw-\om$ contour derived from a
particular $D_E(z)$ is not significantly changed due to changes in
$D_E(x)$ which leave the best-fit $\dw(\om^t)$ unchanged.

We can illustrate the above formal derivation with an example. 
Consider the $w_X(z)$ given by the solid line 
in Fig.~\ref{Fig:wfiddle} (a), which has $w_X(z)=-2$ for 
$0< z < 0.5$, $w_X(z)=0$ for $0.5 < z < 9$ and $w_X(z)=-1$ at higher
redshifts. We keep $\Omega_{\rm m}=0.3$.
This equation of state gives the same angular diameter distance
to $z=1088$ as the $w_X(z)=-1$ simulation, so the CMB constraints
are unchanged. Since the supernovae are most sensitive to low
redshifts, the effective equation of state seen by the supernovae
is closer to -2. 
The resulting conditional $\bar{w}$ versus $\om$ plot is shown
in Fig.~\ref{Fig:wfiddle}(b). 
The solid lines show the constraints from SNAP-like observations
from a simulation using the $w(z)$ shown in Fig.~\ref{Fig:wfiddle}(a).
If we instead simulate SNAP-type data from a model with a constant
equation of state, fixed to the best fitting $\wsn$ at $\om=0.3$
we obtain the dot-dashed line. By design the two coincide at
$\om=0.3$, and overall they are very similar, as expected if
our approximation is good.

We now modify this jumpy $w_X(z)$ model to find the limiting
case where the upper contours only just cross at $\bar{w}=-1$.
We keep the transition redshifts the same, and vary the high
and low values of $w_X$, maintaining the condition that 
$\wcmb=-1$ (at $\om=0.3$). 
We find that a model with $w_X(z)=-1.1$ ($0<z<0.3$),
and $w_X(z)=-0.9$ ($0.3<z<9$) meets this criterion.
This has $\wsn=-1.03$ and and so for this shape of
$w_X(z)$ the solid line of Fig.~\ref{Fig:wsn-wcmb} should actually
pass through the small cross. This illustrates the effect of different
shapes of $w(z)$ on the result in Fig.~\ref{Fig:wsn-wcmb}.

\section{Conclusions\label{sec:conclusions}}

Throughout we have concentrated on the question of 
super-acceleration. However our method can be extended to other
similar questions. For example it can answer the question of whether
or not $w(z)$ ever went \emph{above} $-1$. If the \emph{lower}
contours on the conditional  $\bar{w}$ versus $\om$ plot cross above
$-1$ then we can  be sure that at some point $w(z)$ was greater than
$-1$. Alternatively we can answer questions about different values of
$w$. Denoting the $\bar{w}$ value where the contours  cross as
$\bar{w}_{\rm u}$ (upper contours) and  $\bar{w}_{\rm l}$ (lower
contours) we can say that at some point in the history of the universe
$w(z)<\bar{w}_{\rm u}$, and at some point in the history of the
universe (not necessarily the same point!)
$w(z)>\bar{w}_{\rm l}$. In fact the line in the $\bar{w}$, $\om$ plane
need not even be horizontal. For example, one could imagine testing
for whether  $\bar w<-1/[3(1-\om)]$.  

It is possible that this method could be extended to other methods of
constraining the equation of state of dark energy, for example cosmic
shear and cluster counts. This might make it easier to detect a
super-acceleration phase.  

In conclusion, we have discussed a method which can detect the
super-acceleration of the universe without any assumptions about the
evolution of the dark energy density, or about the present matter
density. This example makes it clear that  
despite the intrinsic degeneracies in supernova
and CMB observations and our ignorance about the evolution of dark
energy density, it will be possible to make fundamental and
robust discoveries about the phenomenological nature of dark energy
with upcoming SNAP-like and Planck observations.

\begin{acknowledgments}
We thank the Aspen Center for Physics where this work was begun. We
thank W. Evans, W. Hu, N. Kaloper, E. Linder, T. D. Saini, J. Weller,
and especially  R. Caldwell and A. Lewis for useful discussions. We
thank D. Huterer and E. Linder for their detailed comments and for
pointing out the importance of the calibration parameter for this
analysis. MK was  supported in part by the NSF and NASA grant
NAG5-11098.  
\end{acknowledgments}


\end{document}